\documentclass[fleqn,usenatbib]{mnras}

\usepackage{newtxtext,newtxmath}

\usepackage[T1]{fontenc}

\DeclareRobustCommand{\VAN}[3]{#2}
\let\VANthebibliography\thebibliography
\def\thebibliography{\DeclareRobustCommand{\VAN}[3]{##3}\VANthebibliography}

\usepackage{graphicx}	
\usepackage{amsmath}	
\usepackage{xspace}
\usepackage{siunitx}
\usepackage{orcidlink}
\usepackage{threeparttable}

\newcommand{\ch}{$\chi^{2}$\xspace}

\title[Transmission Spectrum of GJ 357b]{JWST NIRISS Transmission Spectroscopy of the Super-Earth GJ 357b, a Favourable Target for Atmospheric Retention}

\author[Taylor et al.]{
Jake Taylor$^{\orcidlink{0000-0003-4844-9838},1}$\thanks{E-mail: jake.taylor@physics.ox.ac.uk},
Michael Radica$^{\orcidlink{0000-0002-3328-1203},2,3}$\thanks{NSERC Postdoctoral Fellow},
Richard D. Chatterjee$^{\orcidlink{0009-0008-8739-0932},1}$,
Mark Hammond$^{\orcidlink{0000-0002-6893-522X},1}$,
Tobias Meier$^{\orcidlink{0000-0003-4143-8482},1}$,
\newauthor
Suzanne Aigrain$^{\orcidlink{0000-0003-1453-0574},1}$,
Ryan J. MacDonald$^{\orcidlink{0000-0003-4816-3469},4}$,
Loic Albert$^{\orcidlink{0000-0003-0475-9375},3}$,
Björn Benneke$^{\orcidlink{0000-0001-5578-1498},3}$
\newauthor
Louis-Philippe Coulombe$^{\orcidlink{0000-0002-2195-735X},3}$,
Nicolas B.\ Cowan$^{\orcidlink{0000-0001-6129-5699},5,6}$,
Lisa Dang$^{\orcidlink{0000-0003-4987-6591},3}$
René Doyon$^{\orcidlink{0000-0001-5485-4675},3}$
Laura Flagg$^{\orcidlink{0000-0001-6362-0571},7}$,
\newauthor
Doug Johnstone$^{\orcidlink{0000-0002-6773-459X},8,9}$,
Lisa Kaltenegger$^{\orcidlink{0000-0002-0436-1802},10}$,
David Lafreni\`{e}re$^{\orcidlink{0000-0002-6780-4252},3}$,
Stefan Pelletier$^{\orcidlink{0000-0002-8573-805X},11,3}$,
\newauthor
Caroline Piaulet-Ghorayeb$^{\orcidlink{0000-0002-2875-917X},2,3}$\thanks{E. Margaret Burbridge Postdoctoral Fellow},
Jason F. Rowe$^{\orcidlink{0000-0002-5904-1865},12}$,
Pierre-Alexis Roy$^{\orcidlink{0000-0001-6809-3520},3}$.
\\
$^{1}$Department of Physics, University of Oxford, Parks Rd, Oxford, OX1 3PU, UK\\
$^{2}$Department of Astronomy \& Astrophysics, University of Chicago, 5640 South Ellis Avenue, Chicago, IL 60637, USA\\
$^{3}$Trottier Institute for Research on Exoplanets, Université de Montréal, 1375 Avenue Thérèse-Lavoie-Roux, Montréal, QC, H2V 0B3, Canada\\
$^{4}$Department of Astronomy, University of Michigan, 1085 South University Avenue, Ann Arbor, MI 48109, USA\\
$^{5}$Department of Physics, McGill University, 3600 University St, Montreal, QC H3A 2T8, Canada\\
$^{6}$Department of Earth \& Planetary Sciences, McGill University, 3450 University St, Montréal, H3A 2A7, Canada\\
$^{7}$Department of Physics and Astronomy, Johns Hopkins University, Baltimore, MD, 21218, USA\\
$^{8}$NRC Herzberg Astronomy and Astrophysics, 5071 West Saanich Rd, Victoria, BC, V9E 2E7, Canada \\
$^{9}$Department of Physics and Astronomy, University of Victoria, Victoria, BC, V8P 5C2, Canada \\
$^{10}$Astronomy Department, Cornell University, Ithaca, NY 14853, USA\\
$^{11}$Observatoire astronomique de l’Université de Genève, 51 chemin Pegasi 1290 Versoix, Switzerland\\
$^{12}$Department of Physics and Astronomy, Bishops University, 2600 Rue College, Sherbrooke, QC J1M 1Z7, Canada
}

\date{Accepted 2025 May 30. Received 2025 May 28; in original form 2025 January 13}

\pubyear{\the\year{}}

\begin{document}

\label{firstpage}
\pagerange{\pageref{firstpage}--\pageref{lastpage}}
\maketitle

\begin{abstract}
We present a JWST NIRISS/SOSS transmission spectrum of the super-Earth GJ 357\,b: the first atmospheric observation of this exoplanet. Despite missing the first $\sim$40\% of the transit due to using an out-of-date ephemeris, we still recover a transmission spectrum that does not display any clear signs of atmospheric features. We perform a search for Gaussian-shaped absorption features within the data but find that this analysis yields comparable fits to the observations as a flat line. We compare the transmission spectrum to a grid of atmosphere models and reject, to 3-$\sigma$ confidence, atmospheres with metallicities $\lesssim$100$\times$ solar ($\sim$4\,g/mol) with clouds at pressures down to 0.01 bar. We analyse how the retention of a secondary atmosphere on GJ 357\,b may be possible due to its higher escape velocity compared to an Earth-sized planet and the exceptional inactivity of its host star relative to other M2.5V stars. The star's XUV luminosity decays below the threshold for rapid atmospheric escape early enough that the volcanic revival of an atmosphere of several bars of CO$_2$ is plausible, though subject to considerable uncertainty. Finally, we model the feasibility of detecting an atmosphere on GJ 357\,b with MIRI/LRS, MIRI photometry, and NIRSpec/G395H. We find that, with two eclipses, it would be possible to detect features indicative of an atmosphere or surface. Further to this, with 3--4 transits, it would be possible to detect a 1 bar nitrogen-rich atmosphere with 1000\,ppm of CO$_2$. 
\end{abstract}




\section{Introduction}
\label{sec: Introduction}

The search for atmospheres around terrestrial exoplanets is a key science goal of the James Webb Space Telescope (JWST). To date, there have been an array of observations, in both transmission and emission, attempting to detect atmospheres around terrestrial worlds, but none have yet been fruitful. To first order, the key limitation with determining the presence of an atmosphere around a terrestrial planet is the small size of the planet versus the star. For larger stars, like the Sun, terrestrial planets will have a smaller planet-to-star radius ratio, and therefore a smaller observed transit depth, compared to those orbiting an M-dwarf. This has led to the emergence of the M-dwarf opportunity \citep{dressing_occurrence_2015} --- that terrestrial planets around M-dwarfs are more favorable for atmosphere detection. Moreover, due to their lower luminosity compared to Sun-like stars, the habitable zones around M-dwarfs are located at shorter orbital periods. This makes habitable zone planets around M-dwarfs more likely to transit, and more amenable to repeated atmosphere observations.  


Despite the above benefits for potential detections of terrestrial planet atmospheres, M-dwarf stars also pose significant challenges for this endeavour. Interactions between a star's coronal plasma and dynamo-induced magnetic field generates ionising radiation. Since low-mass stars spin down more slowly than Sun-like ones, their ratio of ionising to bolometric luminosity remains at a maximum for much longer, up to billions of years, and generally remains higher at all ages \citep{JohnstoneFGKM}. The raised fluxes of ionising radiation to which M-dwarfs expose their companion rocky planets can be effective in evaporating primordial hydrogen \citep[e.g.][]{Ho2024} and the volcanic atmospheres that may follow \citep[e.g.][]{chatterjee2024}. Thus, the sweet spot for observing rocky planet atmospheres is around relatively inactive M-dwarfs, which generally means those that started rotating relatively slowly at birth.  

Furthermore, the generally high activity levels of M-dwarfs during their main sequence phase leads to high occurrence rates of photosphere heterogeneities (i.e., spots and faculae on the stellar surface). If occulted during a planet transit, these heterogeneities can contaminate observed light curves \citep[e.g.,][]{fournier-tondreau_near-infrared_2024}. Though, the more problematic scenario is when heterogeneities remain unocculted, in which case they can directly bias transmission spectra themselves via the transit light source (TLS) effect \citep{pont_detection_2008,rackham_transit_2018, rackham_transit_2019, lim_atmospheric_2023, cadieux_transmission_2024, radica_promise_2025}. Moreover, particularly for late-type M-dwarfs, photosphere temperatures are low enough that H$_2$O can condense in their star spots. This can result in H$_2$O spectral features being imprinted on an orbiting planet's transmission spectrum via the TLS effect --- potentially spoofing the presence of an atmosphere on the planet itself. 

Theoretical models have predicted that short-period terrestrial planets with radii $\lesssim$1.4\,R$_\oplus$ should not retain atmospheres that are rich in H$_2$/He \citep{rogers_photoevaporation_2021}. Evidence of a lack of H/He dominated atmospheres around rocky planets was mounting already in the pre-JWST era. For example, observations of the TRAPPIST-1 system \citep[e.g.][]{de_wit_combined_2016,de_wit_atmospheric_2018, wakeford_disentangling_2019,zhou_hubble_2023,garcia_hstwfc3_2022} and the L98-59 system \citep[e.g.][]{damiano_transmission_2022,zhou_hubble_2023} were able to rule out H/He rich atmospheres; however these observations did not have the precision to distinguish between high-mean-molecular-weight atmospheres, clouds, or airlessness. Now, with the unprecedented precision and wavelength coverage of JWST, a new era has begun for the detection and characterization of atmospheres around small, rocky planets.

The exoplanet community commonly uses the cosmic shoreline as a rule of thumb \citep{zahnle_cosmic_2017, redfield_report_2024}--- an empirically motivated boundary distinguishing planets with and without atmospheres based on their instellation and escape velocity. However, the photoevaporation cosmic shoreline does not differentiate between primary and secondary atmospheres or between early and main-sequence escape, implying that planets traditionally classified as airless may, in fact, have evolved to retain atmospheres. One such evolutionary pathway is the volcanic revival of an atmosphere from a bare rock, a scenario that has received little attention since \cite{KiteBarnett}. Stellar ionizing luminosity typically declines more rapidly than exoplanetary mantles are thought to deplete and determining whether volatile supply can then outpace atmospheric loss depends on the nonlinear physics of escape, mantle evolution, and stellar variability \citep[e.g.][]{Nakayama_2022,Dorn18, France_2020}.

\subsection{Overview of JWST Transit Observations of Rocky Planets}
\label{Overview}

As outlined above, probing for atmospheres of terrestrial planets has been one of the main science objectives of JWST over the first few cycles, and numerous such programs have been carried out, both via transit and eclipse observations. Transit programs using NIRSpec/G395H (3--5\,µm), primarily aiming to detect the signatures of CO$_2$ at 4.3\,µm have been published for six planets: TOI-836\,b \citep{alderson_jwst_2024}, L~98-59\,c \citep{scarsdale_jwst_2024}, L~168-9\,c \citep{alam_jwst_2024}, LHS 457\,b \citep{lustig-yaeger_jwst_2023}, GJ 486\,b \citep{moran_high_2023}, and GJ 1132\,b \citep{may_double_2023}. All of these analyses were able to rule out cloud-free, H$_2$/He-dominated atmospheres, but were unable to to definitively rule on the presence of a secondary atmosphere due to the flatness of the observed transmission spectrum \citep{lustig-yaeger_jwst_2023, alderson_jwst_2024, scarsdale_jwst_2024, alam_jwst_2024}, or the degenerate effects of TLS \citep{moran_high_2023, may_double_2023}. However, followup MIRI/LRS (5--12\,µm) eclipse observations of GJ 436\,b suggest that the planet is indeed airless \citep{mansfield_no_2024}.

Transit observations of GJ 341\,b with NIRCam/F444W (3.9--5\,µm) by \citet{kirk_jwstnircam_2024} also yield a similar conclusion. While NIRCam transit spectra of L 98-59\,d by \citet{gressier_hints_2024} and \citet{banerjee_atmospheric_2024} yield tentative evidence of atmospheric features --- the significance of these depended heavily on the particular data reduction and light curve fitting. 

Finally, the two inner planets of the TRAPPIST-1 system \citep{gillon_temperate_2016, gillon_seven_2017} have been observed in transit with NIRISS/SOSS (0.6--2.8\,µm) and in eclipse using MIRI photometry by \citet{lim_atmospheric_2023} and \citet{greene_thermal_2023} respectively for planet b, and \citet{radica_promise_2025} and \citet{zieba_no_2023} respectively for planet c. In both cases, the NIRISS/SOSS observations were dominated by stellar contamination, whereas the MIRI eclipses were consistent with the planets being airless bodies. It should be noted, however, that there are degeneracies in mid-infrared eclipse observations between different potential atmosphere and surface compositions \citep{mansfield_identifying_2019, koll_identifying_2019, ih_constraining_2023, mansfield_no_2024,hammond_reliable_2024} as well as between different thin atmosphere scenarios themselves \citep{zieba_no_2023, lincowski_potential_2023, ih_constraining_2023}.

In this paper, we present a NIRISS/SOSS transmission spectrum of GJ 357\,b, a super-Earth (R$_p$ = 1.217 $\pm$ 0.084 $R_\oplus$, $T_\text{eq}$ = 525 $\pm$ 11 K (assuming zero Bond albedo), P = 3.93 d and M$_p$ = 1.84 $\pm$ 0.31$M_\oplus$) transiting a nearby M2.5V star \citep{luque_planetary_2019,jenkins_gj_2019}. GJ 357 is a slow rotating mid-type M-star with a period of $78 \pm 2$ days \citep{luque_planetary_2019}, which was observed with HARPS to have an activity indicator $\log R_{H K}^{\prime} = 5.53$ \citep{Marvin5.53} based on calcium line emission, making it one of the least active M-dwarfs in a survey of more than 4000 stars \citep{Boro2018}. 
For the rest of this work, we adopt the system parameters of \citet{luque_planetary_2019} as they are generally consistent with, but more precise than those of \citet{jenkins_gj_2019} on account of including additional precise CARMENES and PFS RVs which were not considered in the latter study.

The layout of this paper is as follows: In Section~\ref{sec: Observations} we present the description of our observations and outline our methodology for the data reduction and analysis. In Section \ref{Section3} we present our analysis of the planet's transmission spectrum including atmospheric and internal modelling. In Section \ref{sec: escape} we present modelling that constrains the potential for GJ 357\,b to have retained an atmosphere in the context of the evolution of the host star's ionising luminosity. Then, in Section \ref{future} we describe the potential for future observations of the planet, before concluding in Section \ref{conclusion}.

\section{Observations and Data Analysis}
\label{sec: Observations}

We observed one transit of GJ 357\,b with JWST using the Near Infrared Imager and Slitless Spectrograph (NIRISS) instrument \citep{doyon_near_2023} in Single Object Slitless Spectroscopy (SOSS) mode \citep{albert_near_2023}. The observation began on the 22$^{\text{nd}}$ of November 2023 at 11:40:30 UT and finished at 17:10:32 UT. As the host star is relatively bright ($J$=7.3), we used the SUBSTRIP96 subarray, which only includes the first SOSS diffraction order covering wavelengths 0.85 -- 2.85\,µm. The time series consists of 2201 total integrations with two groups per integration. This observation was part of the NIRISS Exploration of the Atmospheric diversity of Transiting exoplanets (NEAT) GTO program (ID 1201 PI: Lafrenière). 

\begin{figure*}
    \centering
    \includegraphics[width=0.9\textwidth]{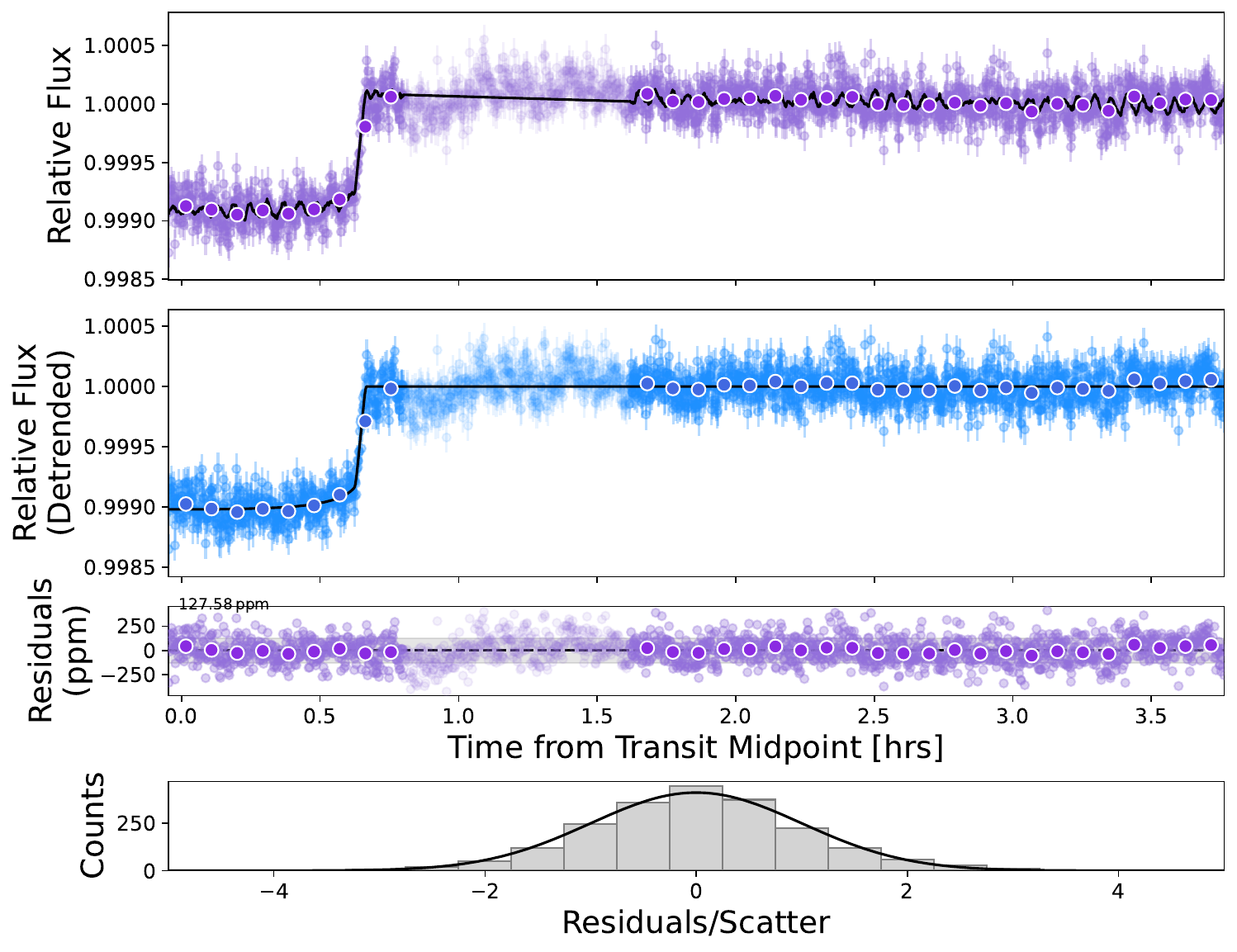}
    \caption{GJ 357\,b NIRISS white light curve fit results.
    \emph{Top panel}: Raw GJ 357\,b order 1 white light curve, with the best-fitting astrophysical transit + systematics model overplotted in black. The first $\sim$40\% of the transit was missed by our observation. Integrations from $\sim$0.8 -- 1.5\,hr post-transit mid-point (shown as faded) were excluded from the fit due to uncorrectable systematics.
    \emph{Second panel}: Systematics corrected white light curve, with the best-fitting astrophysical model overplotted in black.
    \emph{Third panel}: Residuals to the light curve fit. Some correlated noise remains in the post-transit baseline which is not adequately captured by our systematics model. 
    \emph{Bottom panel}: Histogram of residuals.}
    \label{fig: White Light}
\end{figure*}

\subsection{Data Reduction}
\label{sec: Reduction}

We reduced the GJ 357\,b time series observations using the \texttt{exoTEDRF} pipeline \citep{feinstein_early_2023, radica_awesome_2023, radica_exotedrf_2024}, which has been widely employed to analyze both NIRISS \citep{coulombe_broadband_2023, lim_atmospheric_2023, fournier-tondreau_near-infrared_2024, radica_muted_2024, radica_promise_2025, cadieux_transmission_2024, piaulet-ghorayeb_jwstniriss_2024} and NIRSpec observations (\citealp{benneke_jwst_2024}, Schmidt et al.~submitted, Ahrer et al.~submitted).

We apply the standard \texttt{exoTEDRF} Stage 1 corrections \citep[e.g.,][]{radica_awesome_2023, radica_muted_2024, piaulet-ghorayeb_jwstniriss_2024} including superbias and non-linearity corrections, a time-domain cosmic-ray flagging routine \citep{radica_muted_2024}, and ramp fitting. We additionally correct the 1/$f$ noise at the group-level (i.e., before ramp fitting) using the \texttt{scale-achromatic} method \citep{radica_awesome_2023, radica_exotedrf_2024}. In Stage 2, we perform the flat field correction and background subtraction using the STScI-provided SUBSTRIP96 background model. Notably, unlike in many other observations which employ the SUBSTRIP256 subarray \citep[e.g.,][]{lim_atmospheric_2023, fournier-tondreau_near-infrared_2024, cadieux_transmission_2024, radica_muted_2024, gressier_jwst-tst_2024}, we do not find that it is necessary to independently scale the background model before and after the background ``step`` at column $\sim$750. As part of the Stage 2 calibrations, we also employ a principal component analysis (PCA) on the 2D detector images to reveal detector-level trends useful for light curve systematic detrending \citep[e.g.,][]{coulombe_broadband_2023, radica_promise_2025}. As in most SOSS observations, our PCA reveals a sub-pixel drift in the trace position, as well as the characteristic beating pattern due to the telescope's thermal control (for more information see e.g., \citealp{coulombe_broadband_2023}). Finally, we extract the stellar spectra using a simple box extraction with an aperture width of 32 pixels as the order self-contamination cannot be accurately modeled for SUBSTRIP96 observations \citep{radica_applesoss_2022}, and, in any case, is predicted to be negligible \citep{darveau-bernier_atoca_2022}.

\subsection{Light Curve Fits}
\label{sec: Light Curve Fits}

We first summed all the stellar flux extracted from the detector to create a high-S/N white light curve. Unfortunately, as can be clearly seen in Figure~\ref{fig: White Light}, our observations started roughly 10 minutes before the mid-transit point of GJ 357\,b (propagating the mid-transit time from \citet{oddo_characterization_2023} to the date of our observations). We tracked this down to discrepancies between the orbital ephemeris presented in \citet{jenkins_gj_2019}, which we used when constructing the JWST APT file for this observation as its quoted uncertainties were the lowest at the time, and the more accurate ephemeris from \citet{luque_planetary_2019}. When propagated forward to the epoch of our observations, the transit times predicted from the \citet{luque_planetary_2019} or more recent \citet{oddo_characterization_2023} ephemerides agree with what we observed. However, using the \citet{jenkins_gj_2019} ephemeris predicts a mid-transit time $\sim$1.5\,hr later. As a result, we only captured $\sim$60\% of the transit, and missed the entirety of ingress.

Nevertheless, we proceed to fit transit models to this light curve, using the flexible fitting package \texttt{juliet} \citep{espinoza_juliet_2019}. Our light curve model consists of a \texttt{BATMAN} transit model \citep{kreidberg_batman_2015}, with six free parameters: the mid-transit time, $T_0$, the planet-to-star radius ratio, $R_p/R_*$, the scaled orbital semi-major axis, $a/R_*$, the planet's impact parameter, $b$, and the two parameters of the quadratic limb darkening law following the \citet{kipping_efficient_2013} parameterization. We additionally assume a circular orbit \citep{oddo_characterization_2023, luque_planetary_2019}.

We first test the light curve fit using wide, uninformative priors on all of the orbital parameters to see to what degree the transit shape can be accurately recovered from our observations --- especially considering that ingress and egress in particular encode important information about the planet's orbital configuration and stellar limb darkening \citep{seager_unique_2003}. With this method, we find what appears to be an adequate fit to the light curve, however, upon closer inspection, the orbital parameters are highly discrepant (often $>$3$\sigma$) with the published values from \citet{oddo_characterization_2023}. The mid-transit time, in particular, is off by $\sim$12$\sigma$ compared to the time expected by propagating the \citet{oddo_characterization_2023} ephemeris (which is the most up-to-date). We then test putting a Gaussian prior on the mid-transit time, with a mean of the expected transit time (BJD=2460271.0481) and width equal to the propagated uncertainty (0.000908\,d) --- based on the ephemeris from \citet{oddo_characterization_2023}. In this case, we recover orbital parameters which are consistent with those of \citet{oddo_characterization_2023} and \citet{luque_planetary_2019} (i.e., $b=0.213\pm0.118$, $a/R_*=22.62\pm0.678$). We therefore, use this approach for our final fits.

There is correlated noise clearly visible in the post-transit baseline of our light curves, which is not modeled by the astrophysical model alone. To deal with this remaining noise, we include a systematics model that consists of a linear trend in time, as well as with the eigenvalue timeseries of the first two components identified by our PCA (i.e., the trace position drift and beating pattern). Including these in the systematics model decreases the residual point-to-point scatter by $\sim$20\%. Finally, we include an error inflation term added in quadrature to the existing flux errors. We use wide, uninformative priors for all components of the systematics model. 

The best fitting transit models are shown along with the white light curves in Figure~\ref{fig: White Light}. Particularly visible in the third panel, there is correlated noise still present in the post-transit baseline which is not adequately captured by our systematics treatment described above. Gaussian processes (GPs) have been commonly employed to fit out unexplained systematics in NIRISS/SOSS light curves \citep[e,g.,][]{lim_atmospheric_2023, radica_muted_2024, gressier_jwst-tst_2024, cadieux_transmission_2024}, particularly of smaller planets. Periodograms can potentially reveal periodicity in residual noise, potentially correlated with the stellar rotation period, which can then inform the choice of GP kernel (as is common in radial velocity analyses; e.g, \citealp{cloutier_characterization_2017, cloutier_confirmation_2019, radica_revisiting_2022}). However, a periodogram of the baseline integrations does not reveal any significant power at a specific periodicity. 

We also tried employing a GP with a critically damped simple harmonic oscillator (SHO) kernel via the \texttt{celerite} package \citep{foreman-mackey_fast_2017}, which has proven effective at modeling stellar granulation noise in both Kepler \citep{kallinger_connection_2014, foreman-mackey_fast_2017, pereira_gaussian_2019}, and JWST \citep{radica_muted_2024,Coulombe2025highlyreflectivewhiteclouds} light curves. Though, since there is no pre-transit baseline to anchor the GP model, its behaviour is essentially unconstrained during the transit itself, which resulted in clearly unphysical light curve models. This same behaviour persisted irrespective of the axis on which the GP was trained (e.g., time, detector PCs, etc.), the prior restrictions we placed on the GP timescale, as well as the chosen GP kernel itself. We therefore elect not to include any GPs in the final light curve model. 

Finally, we note that the worst offending region for residual correlated noise in the baseline seems to be the region from $\sim$0.8--1.5\,hr post-mid-transit. We thus try repeating our fits without any GP models, but cutting these particular integrations. This test does not result in any significant change to either the inferred orbital parameters, or the final transmission spectrum (e.g., Figure~\ref{fig: Spectrum Compare}), indicating that the residual correlated noise does not have a substantial impact on the transmission spectrum. However, in order to be as conservative as possible, we elect to use the transmission spectrum from the fits where these integrations are cut in the remainder of our analysis. 

In order to produce the transmission spectrum itself, we first bin the light curves to a constant resolution of $R$=50. This resolution provides an adequate tradeoff between having sufficient S/N in individual wavelength bins to accurately model correlated noise, while still well-sampling potential atmospheric features that we hope to detect via Gaussian feature or forward model analyses \citep[e.g.,][]{alderson_jwst_2024, scarsdale_jwst_2024, alam_jwst_2024, Alderson2025}. We then fit the same astrophysical + systematics model described above, but we fix the orbital parameters to the best-fitting values from the white light curve fit. Additionally, as coverage of ingress and egress are critical for constraining limb darkening parameters, freely fitting the limb darkening results in broadly unconstrained posteriors and imprecise transit depths. We, thus, put a Gaussian prior on the quadratic limb darkening terms for each bin, with the mean value set to the prediction from the \texttt{ExoTiC-LD} package \citep{Grant2024} using the 3D Stagger grid \citep{magic_stagger-grid_2015}, and widths of 0.2 \citep{patel_empirical_2022}. The fitted limb-darkening values are shown in Figure~\ref{fig:LD}. Our final spectroscopic fits therefore have eight free parameters: the transit depth and limb darkening, as well as the five parameters of the systematics model describing the transit zero point, linear trends with time and the first two PCA eigenvalue time series, and the error inflation term. The final transmission spectrum is shown in Figures~\ref{fig:Gaussian_test} and \ref{fig:rejection}.

We note that despite the above challenges, we still obtain a transmission spectrum that is sufficiently precise to assess the presence of potential atmosphere scenarios on GJ 357\,b. Particularly in the context of ``previous generation'' observations of small planets with the Hubble Space Telescope where important science was still accomplished despite routinely achieving transit depth precisions $>$100\,ppm at $\sim$1.4\,µm \citep{de_wit_combined_2016, wakeford_disentangling_2019, garcia_hstwfc3_2022}. For comparison, we reach a precision of $\sim$40\,ppm at these wavelengths. Nonetheless, in Figure~\ref{fig: Spectrum Compare}, we compare our achieved transit depth precision to what we could have obtained if we had observed the full transit (and assuming the same level of light curve scatter) as well as the photon-noise precision. We find that our transmission spectrum is $\sim$25\% less precise than if the full transit was captured and $\sim$34\% less than the photon noise precision. The transit depth precisions that we reach, therefore, compare well with previous studies aiming to quantify the compositions of rocky planet atmospheres \citep[e.g.,][]{alderson_jwst_2024, alam_jwst_2024, scarsdale_jwst_2024, Alderson2025}.

\section{Transmission Spectrum and Results}
\label{Section3}

To assess the physical contents of the spectrum we perform three sets of tests: a Gaussian feature detection analysis, a stellar contamination (via the TLS effect) exploration, and a comparison with a grid of forward models. We opt to not perform a full Bayesian spectral retrieval which is common to interpret the atmospheres of terrestrial planets \citep[e.g.,][]{lim_atmospheric_2023,radica_promise_2025}. Given the partial transit and lack of identifiable spectral features, we believe this level of analysis is not warranted for these observations.

\subsection{Feature Detection}

\begin{figure*}
    \centering
    \includegraphics[width=0.99\textwidth]{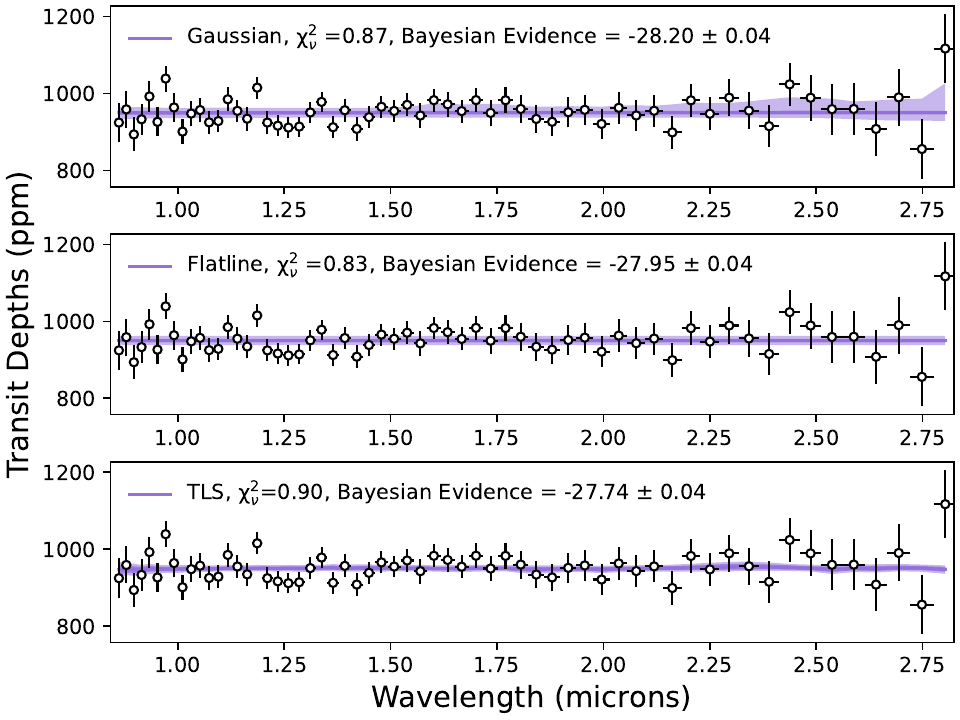}
    \caption{Top panel: Best fitting Gaussian model. Middle panel: flat line model. Bottom panel: Best fitting TLS model. We present the median, 1-$\sigma$, and 3-$\sigma$ contours in purple for the Gaussian model and 1-$\sigma$, and 2-$\sigma$ contours for the flat line and TLS models. It can be seen there is some non-flat structure detected within the 3-$\sigma$ contour of the Gaussian model, however, this model is not favoured over the flat line or TLS model.}
    \label{fig:Gaussian_test}
\end{figure*}

We initially explore whether the data supports any deviations from a flat line, namely any Gaussian features, which could be analogous, and roughly approximate, the broadband features of H$_2$O, CH$_4$ or CO$_2$ \citep[e.g.,][]{jwst_transiting_exoplanet_community_early_release_science_team_identification_2023,moran_high_2023,may_double_2023,scarsdale_jwst_2024}. 

Following the previous work, we parameterize a spectral feature, while remaining agnostic to the particular gas causing the feature, with a Gaussian function:
\begin{equation}
    f(\lambda,A,\mu,\sigma_m,b) = A\,\text{exp}\Bigg(-\frac{(\lambda-\mu)^2)}{2\sigma^2_m}\Bigg) + b
\end{equation}
where $\mu$, $\sigma_m$, and $A$ are the center, standard deviation, and amplitude of the Gaussian respectively. The parameter b is a vertical offset. We fit this model to our data and present the results in Figure~\ref{fig:Gaussian_test} along with the results for a flat line fit --- i.e., using a model of the form:
\begin{equation}
    f(\lambda,b) = b\text{.}
\end{equation}
We set our priors for the amplitude of the Gaussian to be [$-$2,3] in log-space in units ppm, the center of the Gaussian [0.6,3.0] in uniform space and units of microns, the standard deviation of the Gaussian to be [$-$2,0.3]\footnote{The prior choice was made as the lower limit is not smaller than our bin-width, and the upper limit is larger than the feature shape if present.} in log-space with unit microns, and the offset vertical position of the spectrum baseline to be [800,1200] in units ppm. The same priors for the offset are used for our flat-line model. 
We use the nested sampling algorithm \texttt{pymultinest} \citep{buchner_x-ray_2014} with 2000 live points and an evidence tolerance of 0.5\footnote{The evidence tolerance and sampling efficiency are the defaults for \texttt{pymultinest} and recommended for use by the developer.}. We then compare the Bayesian evidence produced from the Gaussian model and flat line model to determine if any features are detected within the data.

We present the best-fitting range of Gaussian models in Figure \ref{fig:Gaussian_test}. We find a Bayesian evidence (ln(Z)) of $-$28.20$\pm$0.04 compared to $-$27.95$\pm$0.04 for the flat line model. We compare these values using the methodology of \citet{trotta_bayes_2008}, whereby the Bayes factor = ln(Z$_1$) $-$ ln(Z$_2$). In our case, Z$_2$ is our flat line model. The larger number indicates that that model is supported more by the data. Thus, we find that the Bayes factor is smaller than 1, and therefore there is inconclusive evidence for any features. In fact, given the slightly larger Bayesian evidence from the flat line model, the data slightly supports the flat line. 

Next, we conducted an array of more physically motivated feature searches, by fixing the position of the Gaussian bump to the wavelengths of known water and methane band heads. Namely for water: 0.95$\mu$m, 1.15$\mu$m, 1.4$\mu$m and 1.85$\mu$m, and for methane: 1.7$\mu$m and 2.3$\mu$m. For water, we find the following Bayesian evidences: 0.95$\mu$m ($-$28.35$\pm$ 0.04), 1.15$\mu$m ($-$28.44$\pm$ 0.04), 1.4$\mu$m ($-$28.39$\pm$ 0.04) and 1.85$\mu$m ($-$28.32$\pm$ 0.04). For methane, we find the following Bayesian evidences: 1.7$\mu$m ($-$28.11$\pm$ 0.04) and 2.3$\mu$m ($-$28.23$\pm$ 0.04). All of these models are less favoured compared to the flat line model. Lastly, we perform the same test at 2.8$\mu$m, as this corresponds to a CO$_2$ band, we find a Bayesian evidence of $-$28.03 $\pm$ 0.04. Therefore, even with our informed prior, we still conclude there to be no features detected within the data. These tests have become important to conduct for observations with low signal-to-noise \citep{Taylor2025}.

\subsection{Transit Light Source Effect Modelling}

As stated in Section \ref{Overview}, previous observations have concluded that GJ 357 is an inactive star, thus we should not see any impacts on the resultant transmission spectrum from the star. We test this by performing a suite of TLS fits to our observations. We model the TLS in a similar fashion to the implementation in \texttt{POSEIDON} \citep{macdonald_poseidon_2023}, assuming four different setups for the star:
\begin{enumerate}
    \item A three-parameter model that assumes a photosphere temperature ($T_\text{phot}$) a heterogeneous region temperature ($T_\text{het}$) and the photosphere covering fraction of the heterogeneity ($f_\text{het}$).
    \item The same as above however with log$(g)$ as free parameter.
    \item A five-parameter model that fits for the photosphere temperature ($T_\text{phot}$), spot temperature ($T_\text{spot}$), spot fraction ($f_\text{spot}$), faculae temperature ($T_\text{fac}$), and faculae fraction ($f_\text{fac}$).
    \item The same as above but with two additional parameters, the log$(g)$ for the spot and faculae regions.
\end{enumerate}
We model the stellar spectra using the PHOENIX grid of models \citep{husser_new_2013}, interpolating using the python package \texttt{pysynphot} \citep{stsci_development_team_pysynphot_2013}. We model the atmosphere assuming it is a flat line. We find the Bayesian evidences to be $-$28.19 $\pm$ 0.04, $-$28.48 $\pm$ 0.03, $-$27.74 $\pm$ 0.04, and $-$28.92 $\pm$ 0.04, for models (i), (ii), (iii), and (iv) respectively. They all provide similar fits to the data, with model (iii) providing the best fit, and only slightly favoured compared to the flat line model. The Bayes factor between the TLS model and flat line is less than 1, and therefore there is no conclusive evidence for this more complex model. We can, therefore, conclude that the impact of stellar contamination is minimal. We have plotted the best fitting models from model (iii) in the bottom panel of Figure \ref{fig:Gaussian_test}.

\subsection{Comparison with Atmospheric Forward Models}
\label{sec: atmospheric_models}
Although both previous analyses do not provide strong evidence for the presence of an atmosphere, flat spectra can also be a result of high atmospheric mean molecular weights or high-altitude clouds. In order to assess what we can physically say about a potential atmosphere on GJ 357\,b, we use \texttt{CHIMERA} \citep{line_systematic_2013} which has recently been coupled with \texttt{FastChem2} \citep{stock_fastchem_2022} \citep[see][for implementation]{schlawin_possible_2024} to generate a grid of atmosphere models as a function of metallicity and cloud top pressure. Similar to \citet{alderson_jwst_2024}, we generate our grid from 1--1000$\times$ solar metallicity, log-spaced with 26 grid points and cloud top pressure from 10--10$^{-4}$ bar, log-spaced with five grid points. For the pressure-temperature profile, we opt for an isothermal temperature fixed at the equilibrium temperature of 525\,K, and we assume that the elemental C/O ratio is solar (where the solar value is 0.55 \citep{asplund_chemical_2009}). We include the following opacities in our model: H$_2$O \citep{polyansky_exomol_2018, freedman_gaseous_2014}, CO$_2$ \citep[][]{freedman_gaseous_2014}, CO \citep{rothman_hitemp_2010}, CH$_4$ \citep{rothman_hitemp_2010}, HCN \citep{barber_exomol_2014}  and NH$_3$ \citep{coles_exomol_2019}. We also model the H$_2$-H$_2$ and H$_2$-He collision-induced absorption (CIA) \citep[][]{richard_new_2012}. We restrict the opacity sources to these as they have prominent absorption cross-sections across the NIRISS/SOSS spectral range \citep{feinstein_early_2023, taylor_awesome_2023}. With our grid, we perform a \ch analysis by fitting each model to the transmission spectrum, computing the rejection sigma as a result. 

To compute the confidence to reject the model, we use the following equation\footnote{Taken from Section 7.2.1 in \citet{gregory_bayesian_2005}}:
\begin{equation}
    \sigma = \frac{\chi^2 - \text{DOF}}{\sqrt{2\times\text{DOF}}}
\end{equation}
where DOF is the degrees of freedom. 

\begin{figure*}
    \centering
    \includegraphics[width=0.49\textwidth]{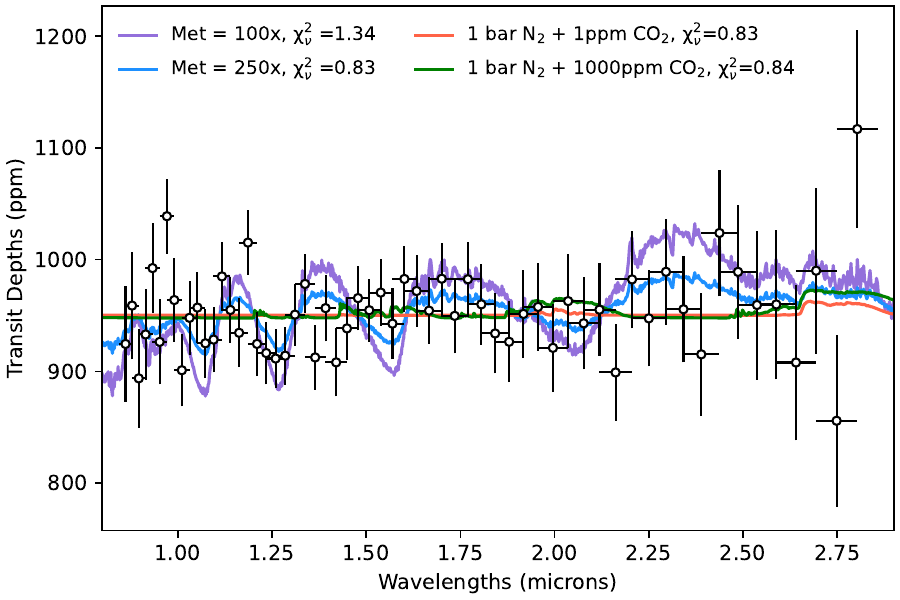}
    \includegraphics[width=0.49\textwidth]{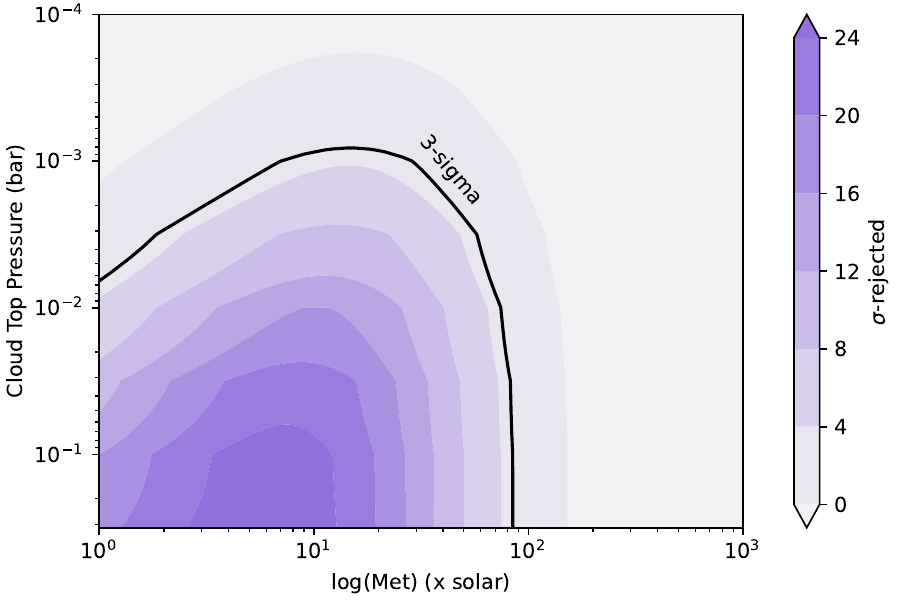}
    \caption{\emph{Left}: Transmission spectrum of GJ 357\,b with four atmospheric models overplotted. The reduced \ch for each model is quoted in the legend. We show two models from our grid analysis: 100$\times$ and 250$\times$ solar metallicity. Each model has an opaque cloud layer at 0.1 bar, we assume this to be analogous to the tropopause of solar system objects \citep{robinson_common_2014}. We also show two nitrogen rich models that would be distinguishable with MIRI, as explained in Section \ref{future}.
    \emph{Right}: Contour plot showing the $\sigma$-rejection of atmospheric models as a function of cloud top pressure and metallicity. The 3-$\sigma$ contour is denoted with the solid black line.}
    \label{fig:rejection}
\end{figure*}

We perform this fitting using \texttt{pymultinest} with one free parameter, the offset between the forward model and the data. We adopt a wide uniform prior for this parameter, as the offsets between the forward model and the data can be large. We show a subset of the models in the left panel of Figure \ref{fig:rejection} and a contour plot displaying the rejection sigma as a function of cloud top pressure and metallicity in right panel. We plot a black line to highlight the 3-$\sigma$ level. Essentially, we can rule out atmospheres that have cloud top pressures higher than around 0.01 bar and metallicities up to around 100$\times$ solar.


In the left panel of Figure \ref{fig:rejection}, we plot a subset of atmospheric models with an associated reduced \ch. We plot the nitrogen rich atmospheres described in Section \ref{future}, alongside models with metallicity of 100$\times$ and 250$\times$ solar. We also consider pure "end-member" atmospheres, specifically pure H$_2$O, CO$_2$, and CH$_4$ and show these in Figure \ref{fig:pure_atm}. In each of these models, we assume a nominal cloud top pressure of 0.1 bar, which is typically where the tropopause is in solar system objects \citep{robinson_common_2014}. For all cases apart from 100$\times$ solar metallicity, the models provide similar fits to the data, with a reduced \ch of less than 1 indicating that we are overfitting, thus we cannot distinguish between these models. Given that we have 60 data points, we would expect a \ch distribution with 59 degrees of freedom and a variance of 118. As we have a sufficiently large number of degrees of freedom, the \ch distribution can be approximated as a Gaussian distribution, therefore, we expect the reduced \ch for our best-fitting model to be drawn from a Gaussian distribution with a mean of unity and a standard deviation of 0.092. Where the standard deviation is taken to be 1 over the root of the variance. The 100$\times$ solar metallicity model, with a reduced \ch of 1.34, can be confidently rejected --- as seen from our rejection tests in the right panel of Figure \ref{fig:rejection}. We also explore the dependence of the C/O ratio and find that this parameter is unconstrained for this quality of data (see Appendix \ref{sec: CtoO_Test}).
\subsection{Interior and Bulk Composition Modelling}
\label{subsec: interior}

\begin{figure}
    \centering
    \includegraphics[width=\linewidth]{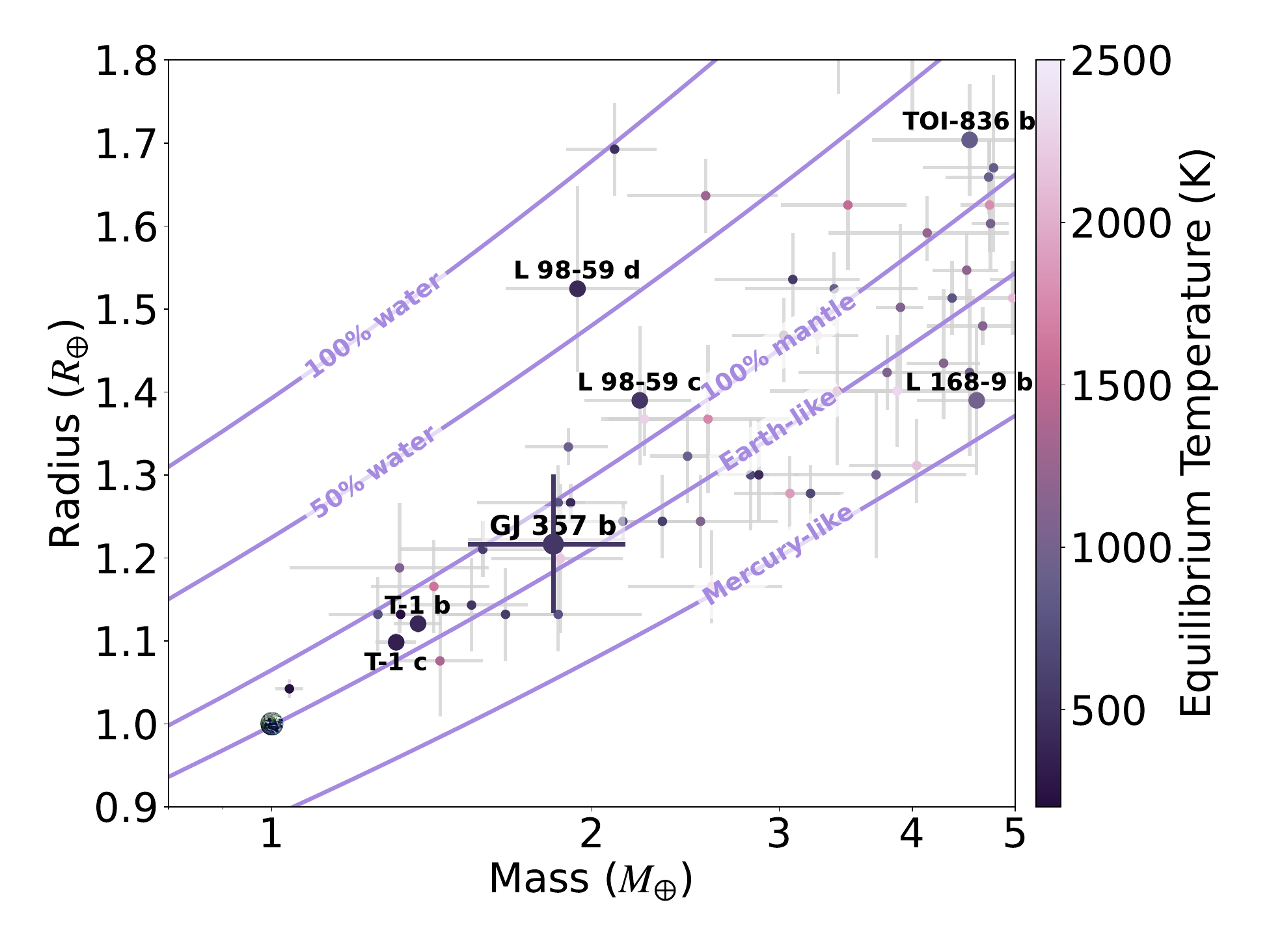}
    \caption{Mass-radius diagram for selected rocky exoplanets retrieved from the DACE database, with planets colour-coded by their equilibrium temperature. The solid curves represent composition models assuming three-layered planetary interiors: a metallic core, a silicate mantle, and a water layer. Planets with published transmission spectra are highlighted in bold. T-1 b and c corresponds to TRAPPIST-1 b and c. GJ 357 b is highlighted as consistent with an Earth-like composition.}
    \label{fig:mr_plot}
\end{figure}

Figure~\ref{fig:mr_plot} shows the mass-radius diagram for potentially rocky exoplanets retrieved from the DACE database. Each planet is colour-coded by its equilibrium temperature. The figure also shows several hypothetical composition curves, which we modelled using the open-source interior structure code MAGRATHEA \citep{Huang2022}. MAGRATHEA solves the hydrostatic equilibrium equations to determine planetary radii for a given mass, based on specified bulk compositions divided into three distinct layers: a metallic core, a silicate mantle, and a water layer. The Earth-like composition assumes 32.5\% Fe and 67.5\% silicates (upper mantle consisting of forsterite, wadsleyite, and ringwoodite and lower mantle consisting of bridgmanite and post-perovskite), representing a metallic core and silicate mantle structure similar to Earth. The 100\%-mantle line corresponds to a planet consisting entirely of MgSiO$_3$, with no core. The Mercury-like curve corresponds to a planet dominated by a metallic core (67\% Fe, 33\% MgSiO$_3$). The 100\%-water line indicates a planet composed entirely of water (in liquid and high-pressure ice form with a surface temperature of $300$\,K), while the 50\%-water line represents a planet with 50\% water layer over an Earth-like mantle and core (16.25 \% Fe and 33.75\% MgSiO$_3$). 

The mass-radius diagram shows that the bulk composition of GJ 357 b is consistent with an Earth-like interior. However, determining the interior structure of a planet based solely on its mass and radius (i.e. its bulk density) is a highly degenerate problem \citep[e.g.][]{Rogers2010, Unterborn2022, Haldemann2024, Guimond2024}. If considering a three-layered planet, at least one of the three layers (core, mantle, hydrosphere) must be fixed to constrain the ratio of the other two layers. Additionally, uncertainties in the composition of each layer, such as the amount of lighter elements in the core \cite[e.g.][]{Hirose2021} or iron content in the mantle \cite[e.g.][]{Unterborn2015, Hakim2018}, further complicate determining the relative proportions of each layer. While GJ 357 b’s bulk composition aligns with an Earth-like interior, it is equally consistent with alternative configurations, such as a higher core-to-mantle fraction including a hydrosphere or a H/He envelope (or both). 
For instance, GJ 357\,b’s mass and radius are equally consistent with a hypothetical interior structure consisting of a pure iron core (with no silicate mantle) and a water layer of approximately $0.54$ Earth masses. Similarly, the planet’s interior bulk composition could be reproduced with an Earth-like silicate mantle and core (core-mass-fraction of $0.325$) and a H/He layer of roughly $0.045$ Earth radii. Measuring the host star’s Mg/Si and Fe/Si elemental ratios has been proposed to help lift this degeneracy \citep{Sotin2007, Dorn2017, Dorn2017a, Spaargaren2023}. 

\section{The Potential for an Atmosphere}
\label{sec: escape}
Efforts to detect rocky exoplanet atmospheres with the James Webb Space Telescope (JWST) are hindered by several factors, including the intense irradiation environments of M-dwarf host stars, which can remove the atmospheres of highly volatile-rich planets \citep{Ji2025}, degeneracies inherent in secondary eclipse observations \citep{hammond_reliable_2024}, and stellar contamination in transmission spectra \citep{radica_promise_2025}. This section discusses why, despite the non-detection of an atmosphere in our observations of GJ 357\,b, its highly inactive M-dwarf host star makes it a relatively strong candidate for follow-up observations aimed at detecting a retained high-MMW atmosphere.

The flat transmission spectrum presented here is consistent with the theoretical expectation of escape of a primordial hydrogen atmosphere from GJ 357\,b. Absorption of X-ray ($1-10$ nm) and EUV ($10-91$ nm) irradiation, altogether XUV, releases photoelectrons that effectively heat the thermosphere and can drive transonic escape, similar to the solar wind predicted by \citet{Parker1958}. The low mass of the host star (0.34 $M_{\sun}$) means GJ 357\,b sees a long and intense saturated phase of ionising radiation, contributing to a large enough lifetime XUV fluence to, in principle, remove a volatile inventory greater than the present-day mass of the planet \citep[e.g.][]{Watson1981, modirrousta-galian_gj_2020}. As such, GJ 357\,b is placed on the airless side of the (XUV) cosmic shoreline \citep{zahnle_cosmic_2017}. However, a high-molecular-weight atmosphere is more resilient to photovaporation and could be outgassed later in the star's evolution--- its presence would also be consistent with our featureless spectrum.

The escape of atmospheres dominated by metals (e.g., C, N, and O) rather than hydrogen is not well explored. Simulations of escape of an Earth-like atmosphere by \cite{Nakayama_2022} found that atomic line cooling prevented rapid loss for extreme fluxes of ionising radiation, up to at least $4 \ \mathrm{W~m^{-2}}$, perhaps precluding bulk escape altogether. However, \cite{chatterjee2024} calculated that a highly ionised Earth-like thermosphere is unstable to escape at lower temperatures ($\lesssim 4500$ K) compared to those assumed by \cite{Nakayama_2022} due to the electrons dragging out ions to space via an ambipolar electrostatic field, effectively reducing mean molecular weight. \citet{chatterjee2024} suggest that at high XUV fluxes the escape of a secondary atmosphere from a super-Earth, such as GJ 357\,b, would take the novel form of a global ion outflow of metals, limited by a thermostat mechanism between photoionisation heating and atomic line cooling. 

\begin{figure}
    \centering
    \includegraphics[width=\linewidth]{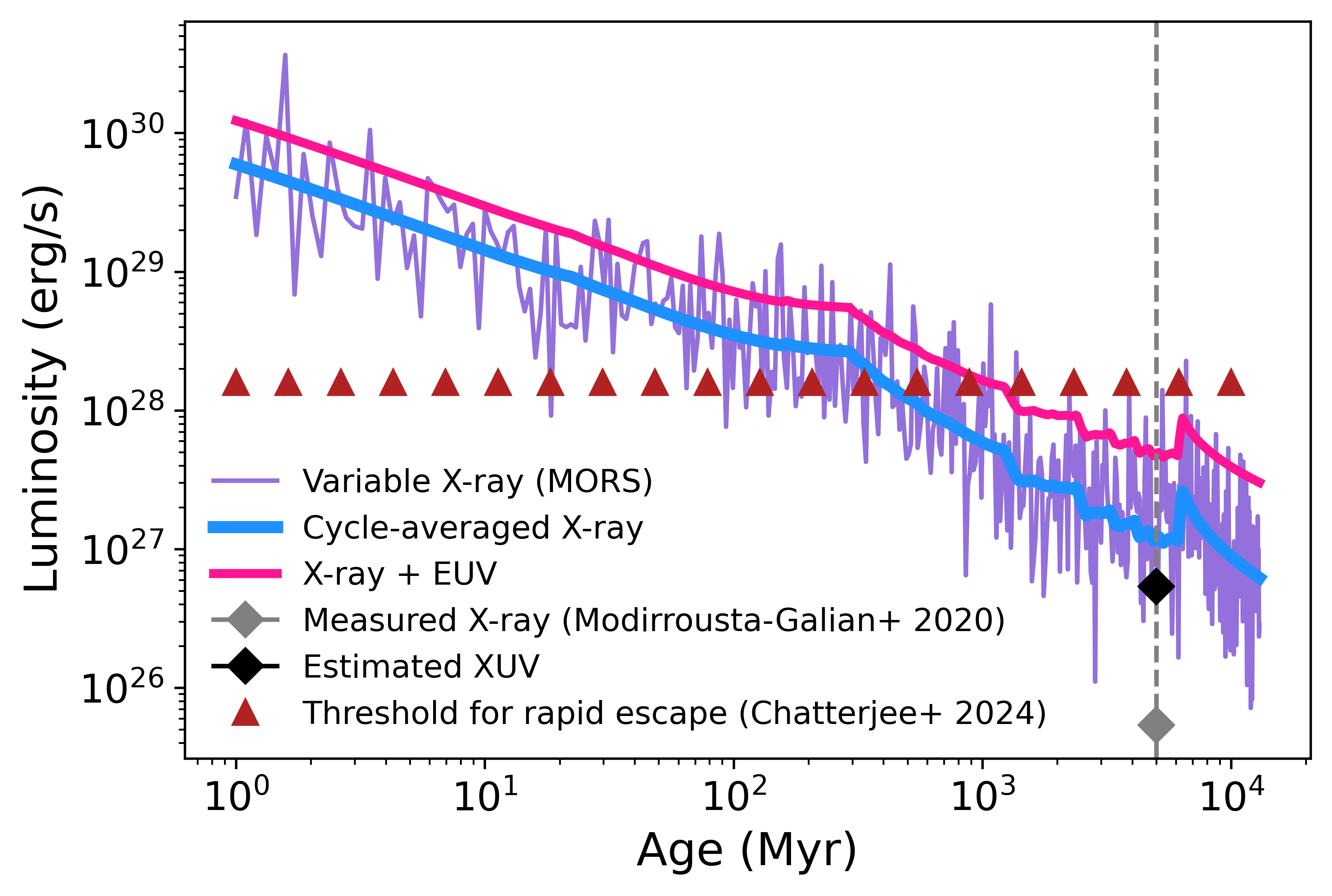}
    \caption{Models of the evolution of the luminosity of GJ 357 in the X-ray and EUV bands. The \textit{XMM-Newton measurement} of the X-luminosity (grey diamond) with the estimated age of 5 Gyrs are taken from \citet{modirrousta-galian_gj_2020}. The XUV luminosity (black diamond) is estimated from the \textit{XMM-Newton} measurement using scaling relationships from \citet{JohnstoneFGKM}. In blue, is the cycle-averaged lowest-activity X-ray track in the evolution of a model cluster of stars of mass $0.34M_{\sun}$ in MORS \citep{JohnstoneFGKM}. The threshold in the XUV luminosity for GJ 357 to drive the hypothetical transonic outflow of a secondary atmosphere on planet b is roughly $10^{28.2} \ \mathrm{erg/s}$. The MORS XUV luminosity (in pink) drops below the threshold for rapid escape \citep{chatterjee2024} at approximately $10^3$ Myrs. \label{fig: XUV}}
\end{figure}

\subsection{Evolution of Ionising Luminosity}

\begin{figure*}
    \centering
    \includegraphics[width=\linewidth]{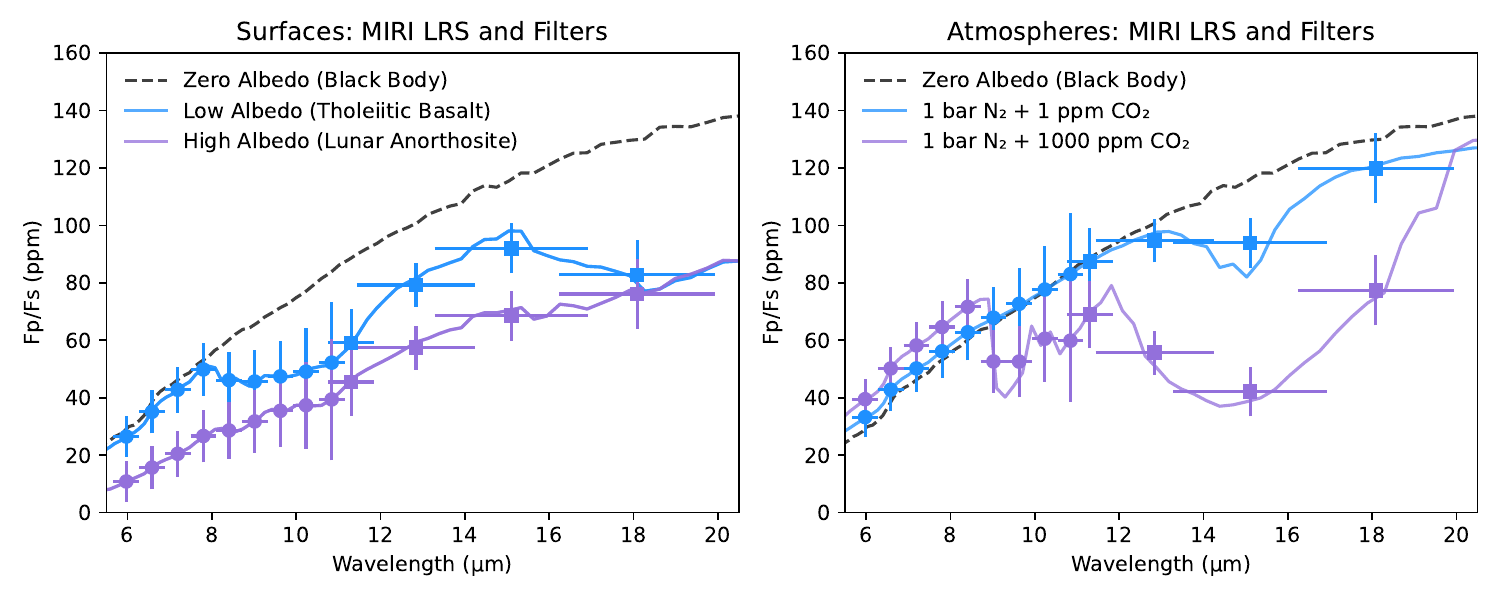}
    \caption{Modelled MIRI LRS and photometric filter emission spectra for GJ 357 b, for simulated surfaces and atmospheres following the methodology of \citet{hammond_reliable_2024}. Left panel: two bare-rock surfaces, with modelled MIRI/LRS (circles) and MIRI filter (squares) emission for two observed eclipses. Right panel: two 1-bar atmospheres, with modelled MIRI/LRS and MIRI filter emission for two observed eclipses. The limited set of cases here are distinguishable from each other at this level of precision given enough spectral coverage, but strong degeneracies exist between observations of individual photometric points.}\label{fig:atmos_surface_miri_emission}
\end{figure*}

To estimate the evolution of the photoevaporation regime of GJ 357\,b we use MORS \citep{JohnstoneFGKM}, a physical rotational evolution model of the high energy emission of stars that is constrained by observed rotational and X-ray distributions in stellar clusters. Direct measurements of the X-ray luminosity are also available with \textit{XMM-Newton} from \cite{modirrousta-galian_gj_2020}, which found $\log L_{\mathrm{x}}\left[\mathrm{erg} \mathrm{s}^{-1}\right]= 25.73 \pm 0.23$ (grey diamond, Fig \ref{fig: XUV}). At such a relatively small X-ray surface flux, the EUV emission is relatively enhanced: we calculate with scaling relations from \cite{JohnstoneFGKM} that the corresponding total XUV luminosity is an order of magnitude higher (black diamond, Fig \ref{fig: XUV}). Figure \ref{fig: XUV} shows that the measured X-ray luminosity is more than an order of magnitude below the main-sequence cycle-average for a star of mass $0.34M_{\sun}$ following the lowest activity track available from MORS \citep[blue, Fig \ref{fig: XUV};][]{JohnstoneFGKM}. The purple track on Figure \ref{fig: XUV} shows the estimated intrinsic variability: the X-ray luminosity will exceed the long-term average by a factor of three roughly 10$\%$ of the evolution phase \citep{JohnstoneFGKM}. Even the troughs of the (lowest activity tracks) variable X-ray evolution are significantly higher than the measured luminosity, showing that GJ 357 b is exceptionally inactive. However, though unlikely for a random sample, it is possible that GJ 357 was observed in an unusually quiescent period and that the long-term average is somewhat higher.

\citet{modirrousta-galian_gj_2020} estimated a minimum age of 5 Gyrs for GJ 357 based on the lower-bound X-ray activity track for a homogeneous grouping of M-dwarfs from \cite{2008Penz}, but this method of estimation does not account for GJ 357 being a mid-M star (M2.5V). We can estimate the age of GJ 357 based on when the rotation period from MORS evolves to that observed of roughly $80$ days, which yields roughly $2.3$ Gyrs. However, given the MORS discrepancy with the measured X-ray luminosity, this estimate is also flawed. Finally, we can compare the observed $\log R_{H K}^{\prime}= 5.53$ activity proxy \citep{Marvin5.53} to trends in an up-to-date M-dwarf sample from \citet{Engle_2024} to estimate a mean minimum age of $4$ Gyrs.

Applying the model from \citet{chatterjee2024} to an upper atmosphere on GJ 357\,b dominated by carbon and oxygen, we find the threshold in ionising radiation to drive transonic escape is $\sim 10^3 F_{\mathrm{xuv}, \earth}$ with a rate of roughly a bar per two million years, where the present XUV flux at Earth is given by $F_{\mathrm{xuv}, \earth}\approx$ \SI{4e-3}{\watt \per \m \squared}. Note that if we instead took the lower bound on the mass of GJ 357\,b of $1.53M_{\earth}$, the threshold would be reduced by less than a factor of 2, which is not enough to change the conclusions made here. Overall, the threshold estimate represents a middle ground between an extrapolation of \cite{Nakayama_2022}, which suggests rapid escape would not occur, and the range of atmospheric heating and dynamics that could enhance escape efficiency \citep[e.g.,][]{Cohen_2024}. Thus, we argue that the possibility of buildup of a secondary atmosphere is limited to when GJ 357\,b is old enough that ionising radiation fluxes have decayed significantly below the best estimate for the threshold for transonic escape (see Figure \ref{fig: XUV}). 

The cycle-averaged XUV luminosity from MORS does not meet the model threshold for bulk escape after one billion years (Fig \ref{fig: XUV}). On taking into account the XUV variability, the MORS track would meet the threshold periodically afterwards. Thus, based on the \textit{model} low-activity track for a star with the same mass as GJ 357\,b, an accumulation of tens of millions of years meeting the threshold could lead to tens of bars of outgassed CO$_2$ being lost, making revival of a secondary atmosphere difficult. However, the \textit{measured} exceptionally low activity means that the cycle-averaged XUV luminosity should have becomes sub-threshold earlier and the steeper time-decay required should severely limit the ability of variability to contribute rapid escape over billions of years. So GJ 357 is significantly more conducive to a retained secondary atmosphere compared to other M2.5V stars.

\subsection{Possible Evolution to a Retained Atmosphere}
In the saturated phase, the loss of a massive primordial hydrogen envelope could strip a planet down to a bare rock. At this stage, volatile escape becomes limited by the rate of outgassing until XUV irradiation decreases below the threshold for rapid photoevaporation. Once this occurs, an atmosphere can gradually accumulate through volcanic activity. Noting that the MORS track significantly overestimates the present X-ray luminosity (Fig \ref{fig: XUV}), we estimate that from $500-2000$ Myr, a growing secondary atmosphere would likely only leak mass to space rather than escape as a hydrodynamic wind.  

The predominant leak of atmospheric mass is likely to be from stellar wind erosion. To estimate this effect, we compare GJ 357 to the similar-sized but significantly more active star GJ 414 ($0.39M_{\sun}$), for which an upper bound on stellar mass loss is constrained to be less than 10\% of the Sun’s rate \citep{Wood_2021}. This upper bound aligns closely with the best estimate for the mass loss rate of TRAPPIST-1 \citep{Dong17}. Accordingly, we infer the atmospheric escape rate by scaling from magnetohydrodynamic simulations of TRAPPIST-1 b’s erosion at 1 bar/Myr \citep{Dong17, Canto1991}, adjusting for the $\sim1/9$ lower wind flux at the $\sim3\times$ greater semi-major axis of GJ 357\,b. Finally, given that stellar wind strength is reduced around less active stars \citep{Wood_2021}, we estimate that, GJ 357\,b’s atmospheric erosion rate would remain well below 0.1 bar/Myr once on the main-sequence.

For a $1.84 M_{\earth}$ stagnant-lid planet \citet{Dorn18} find outgassing rates of $\sim 10$ bar/Gyr for an initial Earth-like mass concentration of radiogenic elements but with considerable sensitivity to various mantle properties. This is similar to the volcanism rate of Earth's midocean ridges, which was adopted by \cite{KiteBarnett} with scaling proportional to mantle depletion. Even though there is significant uncertainty, that the best guess for supply exceeds loss by two orders of magnitude provides some confidence that revival is a distinct possibility. Thus, from the perspective of star-planet interaction and evolution, it is possible that an atmosphere of several to several tens of bars is retained today. 

\subsection{Limitations}
Along with the dynamics of rapid escape of secondary atmospheres being relatively unexplored \citep{chatterjee2024}, there are two major unknowns in this analysis. First, our modelling cannot rule out airlessness due to the possibility that GJ 357 is frequently in a highly energetic flare state \citep[e.g.][]{France_2020, FlaresLowe}. Though flares represent less of a problem for GJ 357\,b than the overall M-dwarf rocky planet population due to the sub-threshold characteristics already discussed. Second, a mantle depleted during the early phase of volatile loss could, in principle, entail later outgassing reduced to $\lesssim 0.1$ bar/Gyr. \citet{krissansen-totton_erosion_2024} modelled how a long-lived magma ocean on TRAPPIST-1\,b enabled catastrophic depletion of the mantle via photoevaporation. However, we note that present-day XUV irradiation of GJ 357\,b is two orders of magnitude lower than at TRAPPIST-1\,b and it would have seen a much shorter and less intense saturated phase, so an extrapolation to similar depletion would be pessimistic.  


\section{Predictions for Future Characterisation}
\label{future}
One key endeavor of the exoplanet community is to detect atmospheres around terrestrial planets, and subsequently, measure their composition. GJ 357\,b is a prime candidate to have retained an atmosphere, as discussed in Section \ref{sec: escape}, therefore, we compute the feasibility of detecting an atmosphere using MIRI. Our analysis is partially motivated by the 500-hour Rocky Worlds DDT program \citep{redfield_report_2024}, which plans to observe this planet as it is well suited to emission spectroscopy.

We model the mid-infrared emission spectra of a small selection of possible atmospheres and surfaces to determine the possibility of measuring their spectral features. We follow the methodology described in \citet{hammond_reliable_2024}, using the \textit{AGNI} model to simulate 1D atmospheres and surfaces with variable albedo and emissivity. From the larger sample modelled in \citet{hammond_reliable_2024}, we chose to model bare-rock surfaces with low and high albedos (tholeiitic basalt and lunar anorthosite), and 1 bar N$_{2}$ atmospheres with 1~ppm and 1000~ppm of CO$_{2}$. These highlight the detectable effects of non-zero albedo, surface spectral features, greenhouse warming of the surface, and the large CO$_{2}$ absorption feature at 15~$\mu m$. We model the observed emission spectra using \texttt{PandExo} \citep{batalha2017pandexo} for MIRI/LRS, and using the instrumental throughputs for the MIRI filters \citep{luger2017planet}.

Figure \ref{fig:atmos_surface_miri_emission} shows the resulting emission spectra, for MIRI/LRS (circles) and for the MIRI F1130W, F1280W, F1500W, and F1800W photometric filters (squares). It shows how variations in the emission spectra will be detectable with just two eclipses, but that strong degeneracies exist between different types of atmosphere and surface, as discussed in \citet{hammond_reliable_2024}. For example, an atmosphere with sufficient CO$_{2}$ could produce notably lower emission particularly around 10~$\mu$m or 15~$\mu$m, which might be distinguishable from the lower levels of spectroscopic variation for surfaces in general \citep{hu2012theoretical}. This could be detected with observations in multiple filters or with MIRI LRS \citep{zieba_no_2023}. Heat redistribution could also lower the total emission from the day-side of an atmosphere by a detectable amount \citep{seager2009method,koll_identifying_2019}, although this may be degenerate with the cooling effect of a high-albedo bare-rock surface \citep{hammond_reliable_2024}. Further observations of transmission spectra may be necessary to break this degeneracy, as phase curve observations may be too time-consuming for this planet given its 3.93-day orbital period. 

\begin{figure*}
    \centering
    \includegraphics[width=0.49\textwidth]{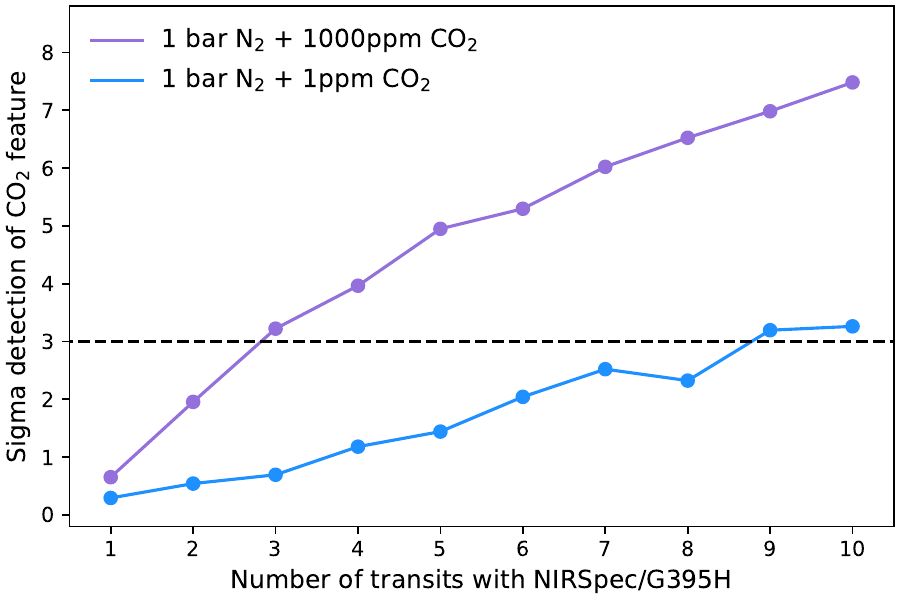}
    \includegraphics[width=0.49\textwidth]{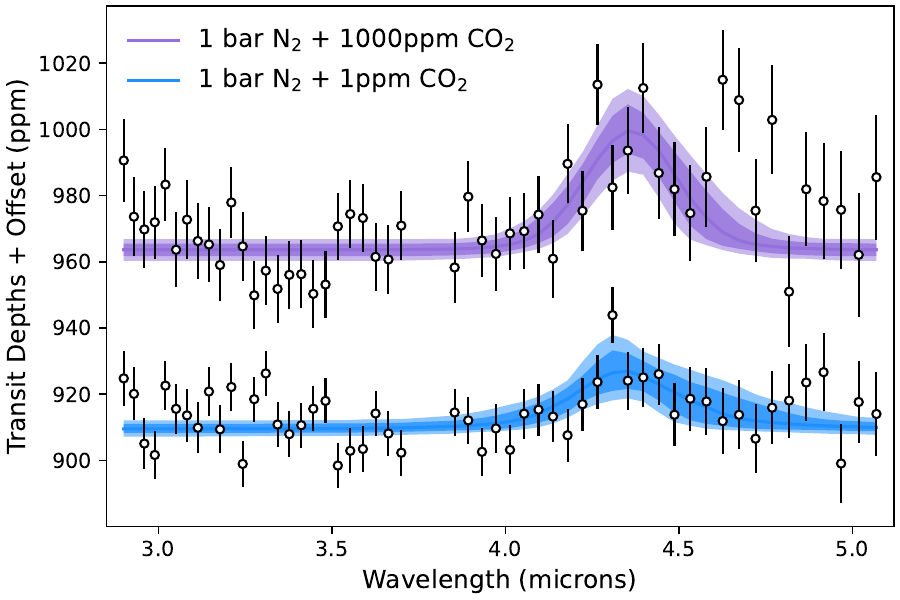}
    \caption{The ability to detect a N$_2$-dominated atmosphere with varying levels of CO$_2$ using NIRSpec/G395H. 
    \emph{Left}: We present the detection significance, based on fitting a Gaussian to the 4.3\,µm CO$_2$ feature, as a function of the number of stacked transits. In purple we show the results from 1 bar N$_2$ atmosphere with 1000\,ppm CO$_2$, and in blue we show the results from 1 bar N$_2$ atmosphere with 1\,ppm CO$_2$. The horizontal black line represents the 3-$\sigma$ threshold, above which can be considered a detection. 
    \emph{Right}: We plot the best fitting models for the lowest number of transits that provides a greater than 3-$\sigma$ detection of the CO$_2$. We add a 50\,ppm offset between the two cases to help visibility.}\label{fig: pandexo_simulations}
\end{figure*}

We explore how many transits would be required to detect the 4.3 micron CO$_2$ feature for both the atmospheres modelled in Figure \ref{fig:atmos_surface_miri_emission}. We take the nitrogen-rich transmission spectra plotted in Figure \ref{fig:rejection} and use \texttt{PandExo} \citep{batalha2017pandexo} to simulate NIRSpec/G395H observations for a resolution of 50. We inject Gaussian noise to the data, centered on the mean value at each wavelength with a standard deviation based on the size of each observational error calculated at each wavelength. We perform our Gaussian feature analysis to determine whether the CO$_2$ feature can be detected and to what sigma can we detect it compared to a flatline model. We do this as a function of transits, assuming that the error on the observations scale as a function of $1/\sqrt{N}$, where N is the number of transits. However, we note that NIRSpec/G395H observations have been shown not to bin down as $1/\sqrt{N}$ \citep[e.g.,][]{alderson_jwst_2024}, so our tests should be treated as an optimistic scenario. For each number of transits, we consider 100 different noise realizations, which we average over to obtain the final detection threshold shown in Figure~\ref{fig: pandexo_simulations}. We find for a 1\,bar N$_2$ atmosphere with 1000\,ppm of CO$_2$, we can detect the CO$_2$ feature to greater than 3-$\sigma$ confidence with 3-4 transits. When decreasing the CO$_2$ abundance to 1\,ppm it takes 9 transits to detect the CO$_2$ feature to $\sim$3-$\sigma$. We plot the sigma detection of CO$_2$ as a function of transit observations, and the corresponding Gaussian feature model, for each scenario in Figure \ref{fig: pandexo_simulations}. We performed the same test for NIRISS/SOSS and found that it is not possible to detect, to 3-$\sigma$, the 1 bar N$_2$ atmosphere with 1ppm CO$_2$. We found that it would take 10 transits to detect the 1 bar N$_2$ atmosphere with 1000ppm CO$_2$ to 3-$\sigma$.

\section{Conclusions}
\label{conclusion}
In this paper, we present the first atmospheric observation of the super-Earth GJ 357\,b. We observed a single transit using NIRISS/SOSS as part of the NEAT GTO program. Despite obtaining only a partial transit (we missed the first $\sim$40$\%$ of the transit), we still recover a transmission spectrum to analyse. 

We explored whether we could detect any spectral features using a Gaussian feature analysis but find that the Gaussian model is disfavored compared to a flat line model with the precision we achieve. We also do not find any statistically significant evidence for stellar contamination in our transmission spectrum despite the late type of the host star. We then model a range of atmospheric scenarios as a function of metallicity and cloud top pressure. We find that we can reject atmospheres lighter than 100$\times$ solar metallicity, with clouds at pressures $>$0.01 bar (see Figure \ref{fig:rejection}). Further observations would be able to determine if heavier atmospheres are present and break the degeneracy between atmospheric metallicity and cloud top pressure. We model the bulk interior composition of the planet using a three-layered model (iron core, silicate mantle, water layer) and find that the interior is consistent with an Earth-like composition. 

We argued that GJ 357\,b is a relatively good candidate for having retained a secondary atmosphere, such as dominated by CO$_2$, which could be elucidated with the NIRSpec/G395H observations from the COMPASS survey. The secondary atmosphere of a cool, $\sim 2M_{\earth}$ super-Earth orbiting an exceptionally quiescent Mid-M star (M2.5V) is expected to be protected from bulk escape on the main sequence \citep{chatterjee2024} and may be optimal for replenishment by volcanism after the active pre-main-sequence phase of volatile loss \citep{Dorn18}. The major uncertainties are the time spent in an energetic flaring state \citep[e.g.][]{France_2020} and the extent of volatile depletion in the mantle on entering the main sequence \citep[e.g.][]{krissansen-totton_erosion_2024}. Furthermore, modelling of evolution scenarios for the GJ 357 would benefit from follow-up X-ray measurements, such as with \textit{XMM-Newton}.

Finally, we predict the ability to detect an atmosphere using MIRI/LRS and/or MIRI photometry. We find that we can determine if the flux deviates from a blackbody with two eclipse measurements with MIRI/LRS or with a number of photometric filters, but these results would be degenerate between a reflective surface and an atmosphere. Transmission spectra can break this degeneracy, therefore, we compute transmission spectra based on these simulations and show that they are consistent with the current observations, we cannot rule them out. We simulated the ability to detect an atmosphere in transmission using NIRSpec/G395H, with a focus on the 4.3\,µm CO$_2$ feature. We find that, with 3-4 transit observations, it would be possible to detect a 1\,bar N$_2$ atmosphere with 1000\,ppm of CO$_2$. Our analysis demonstrates the favourability to observe GJ 357b in both transmission and emission with JWST.

\section*{Acknowledgements}
The authors thank the anonymous reviewer for their helpful feedback which improved the clarity of our analysis and manuscript. The authors thank Darius Modirrousta-Galian and Ignazio Pillitteri for help with interpreting the XMM-Newton observations. We thank Claire Guimond and Raymond Pierrehumbert for feedback on the initial manuscript which greatly improved our clarity.
J.T. was supported by the Glasstone Benefaction, University of Oxford (Violette and Samuel Glasstone Research Fellowships in Science 2024).
M.R.\ would like to acknowledge funding from the Natural Sciences and Research Council of Canada (NSERC), as well as from the Fonds de Recherche du Québec --- Nature et Technologies (FRQNT). This project was undertaken with the financial support of the Canadian Space Agency.
This project has been carried out within the framework of the National Centre of Competence in Research PlanetS supported by the Swiss National Science Foundation under grant 51NF40\_205606. S.P.\ acknowledges the financial support of the SNSF.
L.D. is a Banting and Trottier Postdoctoral Fellow and acknowledges support from NSERC and the Trottier Family Foundation.
D.J.\ is supported by NRC Canada and by an NSERC Discovery Grant.
NBC acknowledges support from an NSERC Discovery Grant, a Tier 2 Canada Research Chair, and an Arthur B.\ McDonald Fellowship. 
The authors also thank the Trottier Space Institute and l’Institut de recherche sur les exoplanètes for their financial support and dynamic intellectual environment.
This work is based on observations made with the NASA/ESA/CSA JWST. R.D.C acknowledges support from the Science and Technology Facilities Council (STFC) and the Alfred P. Sloan Foundation under grant G202114194 (AEThER). The data were obtained from the Mikulski Archive for Space Telescopes at the Space Telescope Science Institute, which is operated by the Association of Universities for Research in Astronomy, Inc., under NASA contract NAS 5-03127 for JWST. The specific observations analyzed can be accessed via DOI \url{10.17909/zq0q-jd03}. 
This research has made use of the NASA Exoplanet Archive, which is operated by the California Institute of Technology, under contract with the National Aeronautics and Space Administration under the Exoplanet Exploration Program.
This publication makes use of The Data \& Analysis Center for Exoplanets (DACE), which is a facility based at the University of Geneva (CH) dedicated to extrasolar planets data visualisation, exchange and analysis. DACE is a platform of the Swiss National Centre of Competence in Research (NCCR) PlanetS, federating the Swiss expertise in Exoplanet research. The DACE platform is available at \url{https://dace.unige.ch}.

\section*{Data Availability}
The data is available upon request.



\bibliographystyle{mnras}
\bibliography{references,bib,refs_tgm,mcr} 



\appendix
\section{Exploring the C/O dependence}
\label{sec: CtoO_Test}
\begin{figure*}
    \centering
    \includegraphics[width=0.99\textwidth]{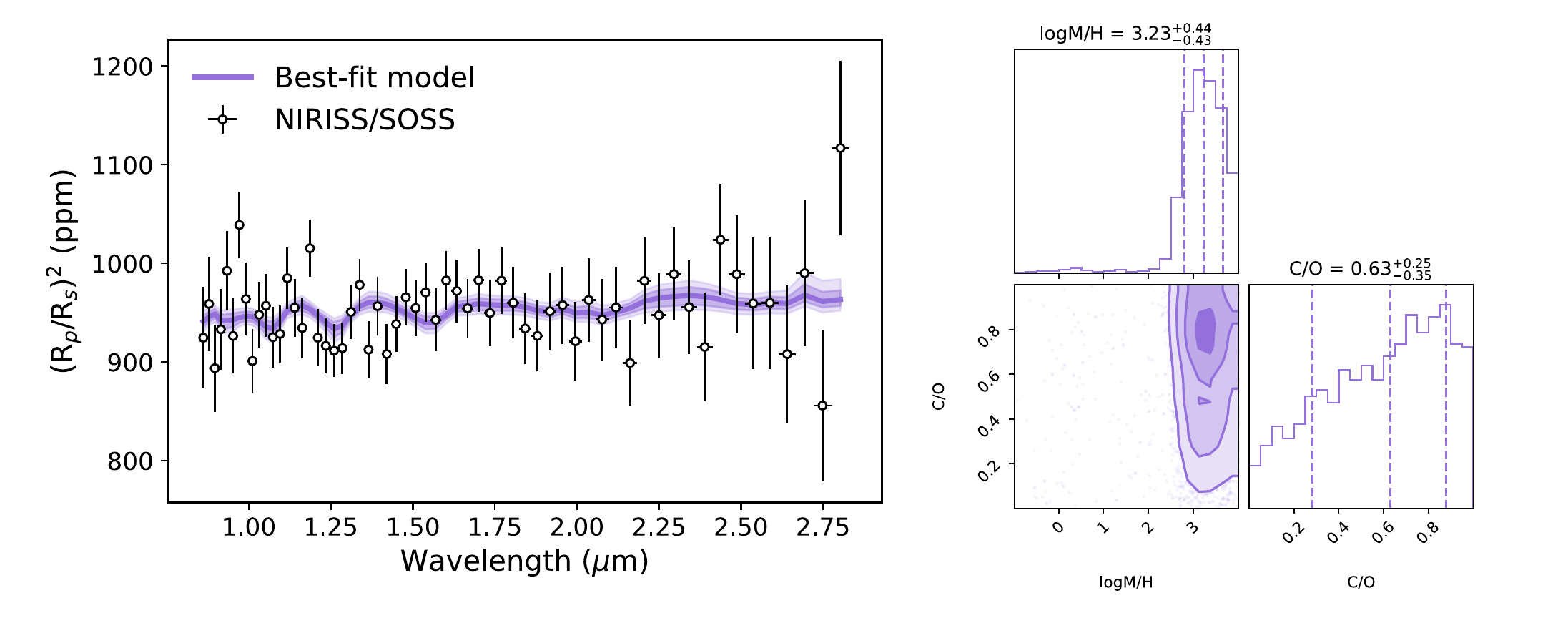}
    \caption{Best-fitting models and posterior distributions of the metallicity and C/O ratio from a simple atmospheric retrieval based on the model described in Section \ref{sec: atmospheric_models}. We show that the atmospheric inferences has no dependence on the C/O ratio. }
    \label{fig:Retrieval_Plot}
\end{figure*}

We perform an atmospheric retrieval with the setup described in Section \ref{sec: atmospheric_models}. We have 5 free parameters: metallicity with priors [-1,4] in log-space, C/O ratio with priors [0,1] in linear-space, planet-radius scaling factor with priors [0.8,1.2] in linear space, cloud top pressure with priors [-6,2] in log-space, and temperature with priors [300,800] in linear-space. We plot the best-fitting range of spectra and posterior distribution of the metallicity versus C/O in Figure \ref{fig:Retrieval_Plot}. We find that there is no dependence on the C/O ratio.

\section{Additional Figures \& Tables}

\begin{figure*}
    \centering
    \includegraphics[width=0.9\textwidth]{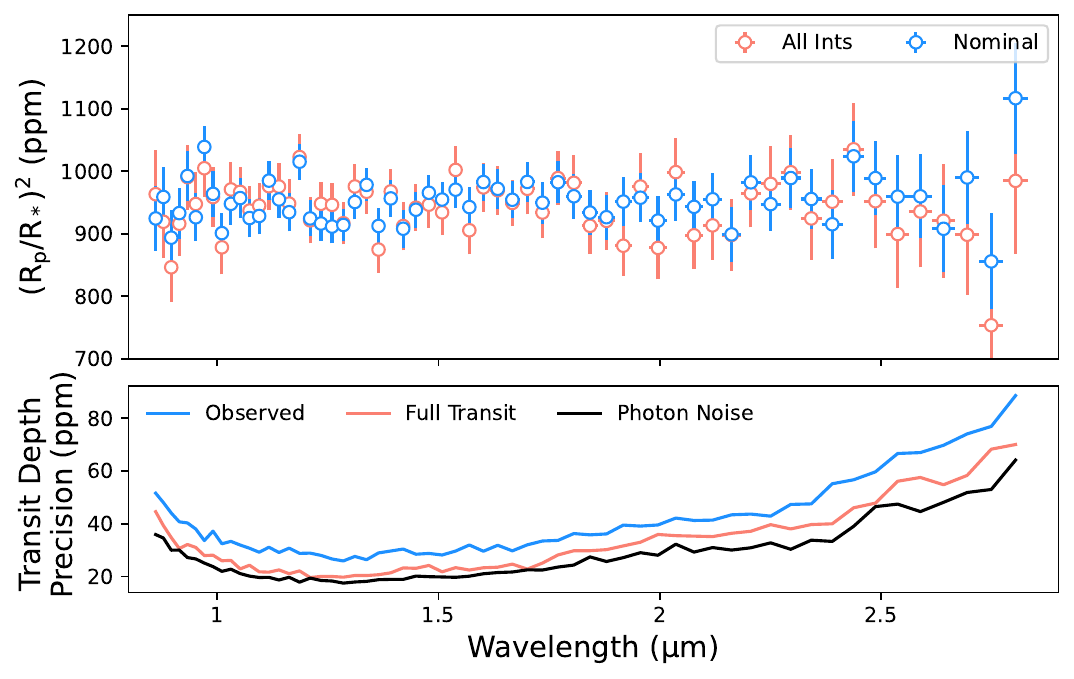}
    \caption{\emph{Top}: Comparison of our nominal NIRISS/SOSS transmission spectrum of GJ 357\,b (blue), obtained cutting the $\sim$500 integrations after transit egress most affected by uncorrectable correlated noise during the light curve fits, and the transmission spectrum obtained by including all integrations in the fit (red).
    \emph{Bottom}: Effects of observing only a partial transit on the achieved transit depth precision. Our observed precision (blue) is compared to the precision that we would have achieved if the full transit was observed (red) and the photon-noise precision (black).}
    \label{fig: Spectrum Compare}
\end{figure*}

Figure~\ref{fig: Spectrum Compare} shows a comparison between the NIRISS/SOSS GJ 357\,b transmission that we obtain with and without cutting integrations affected by correlated noise. 

\begin{figure*}
    \centering
    \includegraphics[width=0.65\textwidth]{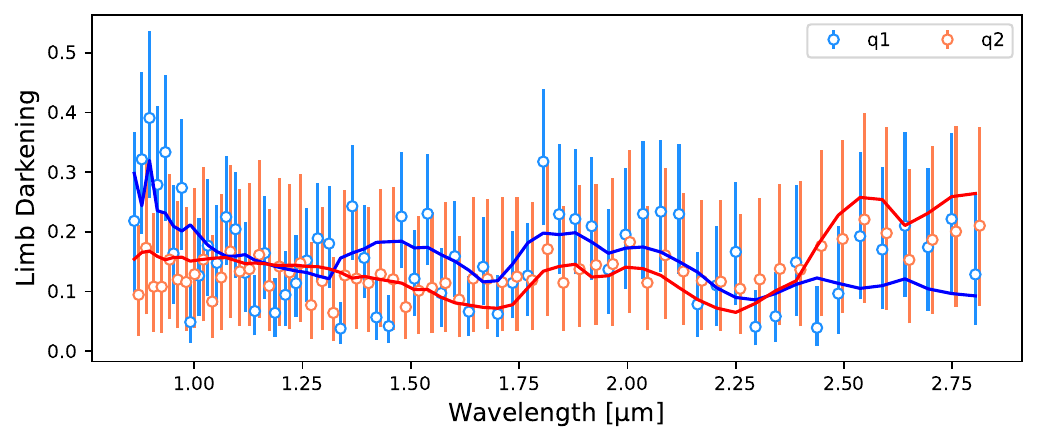}
    \caption{Best-fitting quadratic limb darkening values as a function of wavelength (points with error bars), compared to \texttt{ExoTiC-LD} predictions (solid lines) using 3D Stagger stellar models. Points are slightly offset in wavelength space for visual clarity.}
    \label{fig:LD}
\end{figure*}

Figure~\ref{fig:LD} shows the best-fitting limb darkening values from our spectroscopic fits, compared to stellar models

Table~\ref{tab: spectrum} has the \texttt{exoTEDRF} transmission spectrum used in this work.

\begin{table*}
\centering
\caption{\texttt{exoTEDRF} transmission spectrum of GJ 357\,b used in this work}
\label{tab: spectrum}
\begin{threeparttable}
    \begin{tabular}{cccc}
        \hline
        \hline
        Wavelength & Wavelength Error & Transit Depth & Transit Depth Error \\
        (µm) & (µm) & (ppm) & (ppm)  \\
        \hline
        0.862 & 0.009 & 924.53 & 51.62 \\
        0.879 & 0.009 & 958.87 & 48.08 \\
        0.897 & 0.009 & 893.72 & 44.12 \\
        0.915 & 0.009 & 933.02 & 40.77 \\
        0.933 & 0.009 & 992.42 & 40.35 \\
        0.952 & 0.010 & 926.36 & 38.04 \\
        0.971 & 0.010 & 1038.87 & 33.66 \\
        0.991 & 0.010 & 963.82 & 37.24 \\
        1.011 & 0.010 & 901.00 & 32.46 \\
        1.031 & 0.010 & 947.91 & 33.36 \\
        1.052 & 0.011 & 957.06 & 31.98 \\
        1.074 & 0.011 & 925.11 & 30.75 \\
        1.095 & 0.011 & 928.41 & 29.24 \\
        1.117 & 0.011 & 984.98 & 31.14 \\
        1.140 & 0.011 & 954.93 & 29.15 \\
        1.163 & 0.012 & 934.49 & 30.78 \\
        1.186 & 0.012 & 1015.27 & 28.81 \\
        1.210 & 0.012 & 924.54 & 28.91 \\
        1.235 & 0.012 & 916.39 & 28.04 \\
        1.260 & 0.013 & 911.47 & 26.68 \\
        1.285 & 0.013 & 914.01 & 25.94 \\
        1.311 & 0.013 & 950.78 & 27.66 \\
        1.338 & 0.013 & 978.23 & 26.41 \\
        1.365 & 0.014 & 912.46 & 28.98 \\
        1.392 & 0.014 & 956.57 & 29.71 \\
        1.420 & 0.014 & 908.03 & 30.45 \\
        1.449 & 0.014 & 938.43 & 28.49 \\
        1.478 & 0.015 & 965.56 & 28.80 \\
        1.508 & 0.015 & 954.70 & 28.20 \\
        1.539 & 0.015 & 970.51 & 29.69 \\
        1.570 & 0.016 & 942.70 & 31.99 \\
        1.602 & 0.016 & 982.71 & 29.62 \\
        1.634 & 0.016 & 971.91 & 31.89 \\
        1.667 & 0.017 & 954.35 & 29.77 \\
        1.701 & 0.017 & 982.90 & 32.05 \\
        1.735 & 0.017 & 949.66 & 33.50 \\
        1.770 & 0.018 & 982.45 & 33.73 \\
        1.806 & 0.018 & 959.96 & 36.32 \\
        1.842 & 0.018 & 933.92 & 35.83 \\
        1.879 & 0.019 & 926.47 & 36.18 \\
        1.917 & 0.019 & 951.41 & 39.53 \\
        1.956 & 0.020 & 957.67 & 39.15 \\
        1.996 & 0.020 & 921.03 & 39.58 \\
        2.036 & 0.020 & 962.88 & 42.19 \\
        2.077 & 0.021 & 943.02 & 41.28 \\
        2.119 & 0.021 & 955.20 & 41.41 \\
        2.162 & 0.022 & 899.02 & 43.42 \\
        2.206 & 0.022 & 982.27 & 43.71 \\
        2.250 & 0.023 & 947.40 & 42.92 \\
        2.296 & 0.023 & 989.12 & 47.37 \\
        2.342 & 0.023 & 955.55 & 47.56 \\
        2.389 & 0.024 & 915.21 & 55.19 \\
        2.438 & 0.024 & 1023.87 & 56.71 \\
        2.487 & 0.025 & 989.04 & 59.71 \\
        2.537 & 0.025 & 959.37 & 66.67 \\
        2.588 & 0.026 & 959.90 & 67.01 \\
        2.641 & 0.026 & 907.92 & 69.79 \\
        2.694 & 0.027 & 990.07 & 74.10 \\
        2.748 & 0.027 & 855.77 & 76.93 \\
        2.803 & 0.027 & 1116.87 & 88.60 \\
        \hline
    \end{tabular}
\end{threeparttable}
\end{table*}

\begin{figure*}
    \centering
    \includegraphics[width=0.75\textwidth]{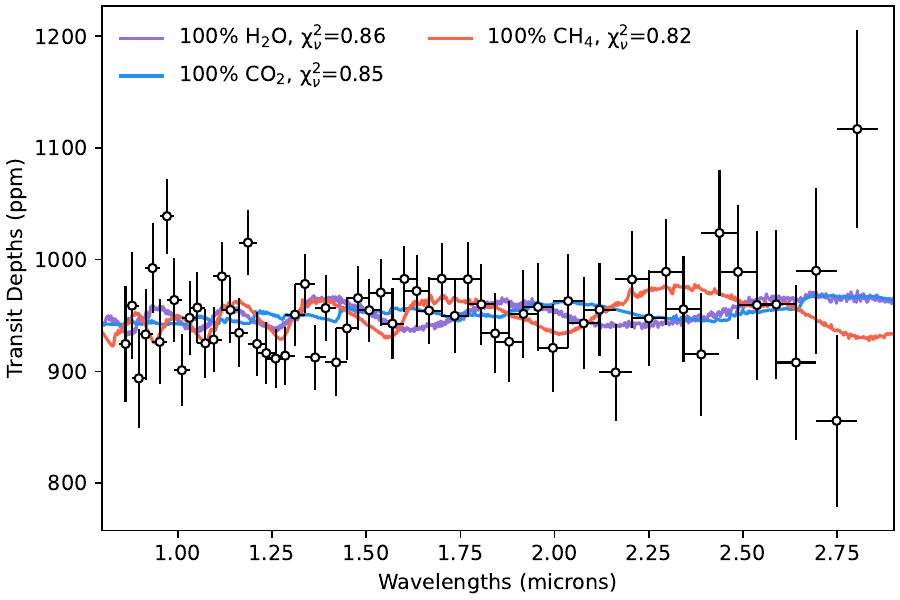}
    \caption{Transmission spectrum of GJ 357\,b with three atmospheric models overplotted. The reduced \ch for each model is quoted in the legend. We show three "end-member" models: pure-H$_2$O, pure-CO$_2$, and pure-CH$_4$.}
    \label{fig:pure_atm}
\end{figure*}

\bsp	
\label{lastpage}
\end{document}